\documentclass[aps,prd,amsmath,twocolumn,longbibliography]{revtex4-1}

\usepackage{graphicx}
\usepackage{amssymb,color}
\usepackage{float}
\usepackage{subcaption}
\usepackage{array}
\newcolumntype{P}[1]{>{\centering\arraybackslash}p{#1}}
\renewcommand{\bar}[1]{\overline{#1}}
\renewcommand{\bar}[1]{\overline{#1}}
\newcommand{\half}{{\frac{1}{2}}}

\def\e{\epsilon}
\def\Dslash{\raise.15ex\hbox{/}\kern-.7em D}
\def\Pslash{\raise.15ex\hbox{/}\kern-.7em P}

  \def\Cv{{\cal V}}

\def\t{\tau}

\newcommand{\ra}{\rangle}

\newcommand{\ben}{\begin{displaymath}}
\newcommand{\een}{\end{displaymath}}
\newcommand{\be}{\begin{equation}}
\newcommand{\ee}{\end{equation}}
\newcommand{\bea}{\begin{eqnarray}}
\newcommand{\eea}{\end{eqnarray}}

\newcommand{\eq}[1]{Eq.~(\ref{#1})}

\newcommand{\bfp}{{\bf p}}

\def\r{\rho}\def\d{\delta}

\def\m{\mu}

\newcommand{\beqn}{\begin{equation}}
\newcommand{\eeqn}{\end{equation}}

 \def\D{\Delta}

\begin{document}

NT@UW-24-04 \title{
Electromagnetic form factors for nucleons in short-range correlations and the EMC effect}

\author{Dmitriy N. Kim$^1$, Or Hen$^2$,  Gerald A. Miller$^1$,  E. Piasetzky$^3$,  M. Strikman$^4$ and L. Weinstein$^5$}

\affiliation{$^1$ 
Department of Physics,
University of Washington, Seattle, WA 98195-1560, USA}

 \affiliation{$^2$
Department of Physics,
Massachusetts Institute of Technology, Cambridge Massachusetts 02139, USA }
        
        \affiliation{$^3$ 
School of Physics and Astronomy, Tel Aviv University, Tel Aviv 69978, Israel}

\affiliation{ $^4$
Department of Physics,
The Pennsylvania State  University, State College, PA 16802  , USA}
\affiliation{ $^5$
Department of Physics,
Old Dominion University,   Norfolk, Virginia, 23529, USA }
                                                                           
\date{\today}     
\begin{abstract}
The relationship between medium modifications of nucleon electromagnetic form factors and nucleon  structure functions is examined using a model motivated by Light-Front Holographic QCD (LFHQCD). These modifications are closely connected with the influence of short-ranged correlations. The size of the modifications to nucleon form factors is shown to be about the same as the modifications to the structure functions. Thus, small limits on form factors modifications do not rule out an
 explanation of the EMC effect motivated by the influence of short range correlations, as claimed by a recent paper.
\end{abstract}
\maketitle

Recent experimental studies~\cite{Weinstein:2010rt,Hen:2016kwk,Duer:2018sxh,Schmookler:2019nvf}   seem to support models in which the EMC effect (medium modification of nucleon structure functions at $0.3 < x < 0.7$) is mainly due to the influence of nucleon-nucleon short-range correlations (SRCs) (high-relative momentum nucleon-nucleon correlations). 
This hypothesis requires that the structure functions for nucleons involved in SRCs be heavily suppressed compared to that of free nucleons.
     
 Ref.~\cite{Xing:2023uhj} claimed that medium modifications of structure functions motivated by the influence of short range  correlations causes a very large strong suppression of elastic electromagnetic form factors of bound nucleons at values of momentum transfer, $Q^2>0.25\, \rm GeV^2$.  Since this strong suppression seems unlikely, Ref.~\cite{Xing:2023uhj} casts 
doubt on the causal nature of the EMC-SRC correlation.   

The expansion of a nucleon's radius in the medium was one of the early proposed hypotheses for the explanation of the EMC effect,   see the reviews~\cite{Geesaman:1995yd,PRNorton_2003,Hen:2013oha,Rith:2014tma,Malace:2014uea,Hen:2016kwk}. Thus the suggestion of Ref.~\cite{Xing:2023uhj} encourages us to seek a deeper understanding of the connections between the medium modifications of nucleon structure functions and elastic form factors. It turns out that in doing so we shall refute the claims of Ref.~\cite{Xing:2023uhj}.
 
There are strong constraints on medium modifications of form factors \textemdash they have been shown to be small~\cite{Sick:1985ygc}, with an upper limit between 3 and 6\% for values of $Q^2$ between 1 and 4 GeV$^2$, for values of missing momentum ($p$ of Fig.~1) less than about 150 MeV/c. Therefore it is necessary to verify that any model of the EMC effect does not violate such constraints. Constraints for larger values of the missing momentum do not seem to exist.
 
The purpose of this is paper is to study  the connection between medium modifications of nucleon structure functions and elastic electromagnetic form factors, as raised in the early papers, Refs.~\cite{Frankfurt:1985cv,CiofidegliAtti:2007ork}. 

We begin with a necessarily somewhat lengthy summary of previous work related to the current effort.
The early papers showed that the nuclear modification of the nucleon wave function depends  on the virtuality, $\mathcal{V}=p^2-M^2$, of the nucleon, where $p$ is the four-momentum of the bound nucleon and $M$ is the nucleon mass.
The relevance of virtuality can be seen from the simple Feynman diagram shown in Fig.~1. If the energy of the $A-1$ residual nucleus is given by $M_A-M+\Delta E$, the use of four-momentum conservation tells us that $p^2-M^2=- 2M \D E+\D E^2$.  If one is examining a regime in which the scattering is from a correlated pair of nucleons (quasi-deuteron model) then $\mathcal{V}\approx -2 \bfp^2$. Small values of $\D E$ correspond to typical mean-field effects, and larger values generally arise from the effects of SRCs. The structure of the nucleon depends on the value of $p^2$ that is necessarily different than $M^2 $.

   \begin{figure}[h]
  \centering
 \includegraphics[width=0.5\textwidth]{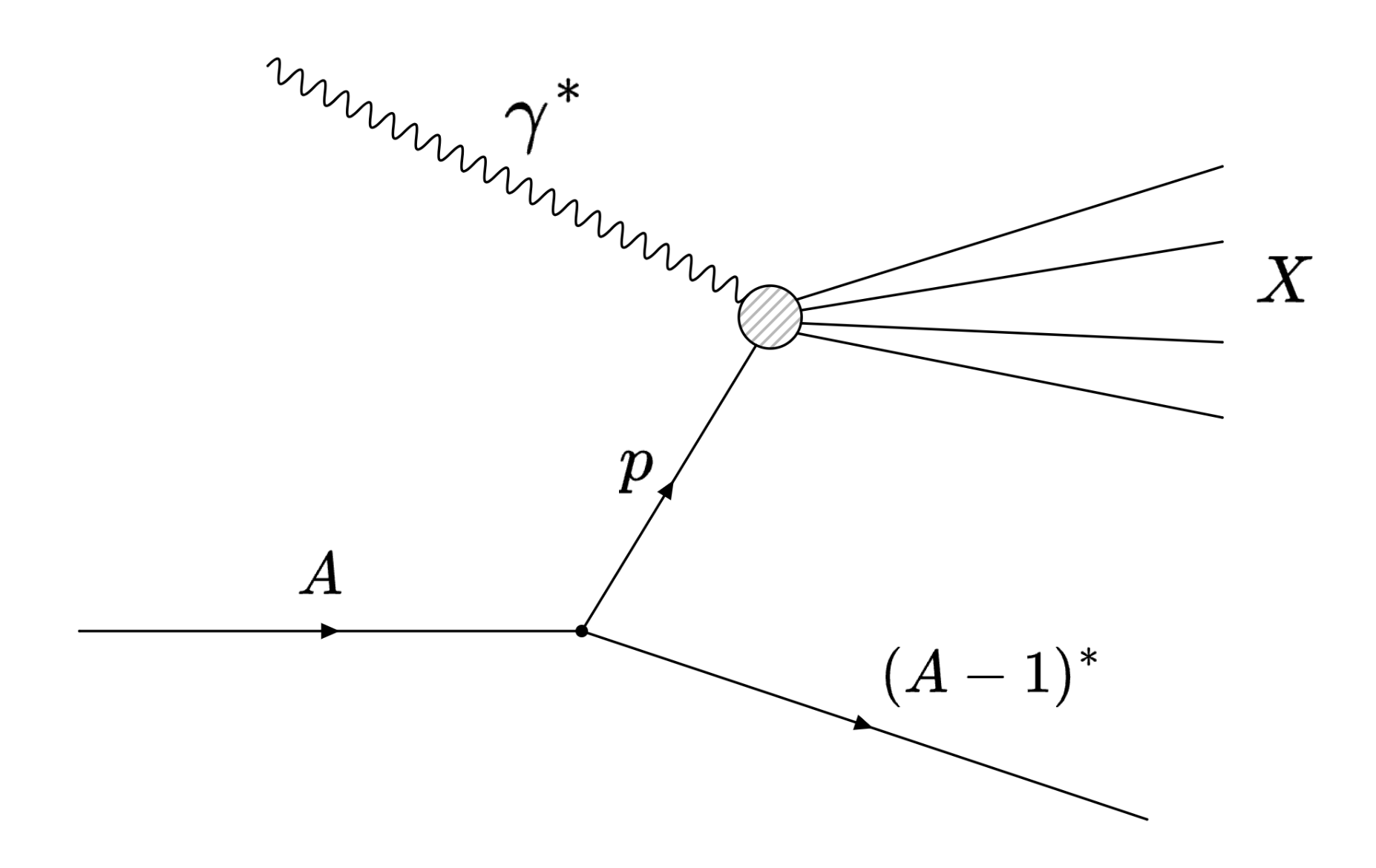}
  \hspace{0.3cm}
  \caption{ Deep inelastic scattering event leaving a residual nucleus $(A-1)^*$ in an excited state.}
\label{DIS}  \end{figure}

It was argued that larger values of the magnitude of the virtuality led to larger suppression of the point-like configurations (PLCs) \cite{Frankfurt:1985cv}, quark configurations that dominate the high-$x$ behavior of nucleon structure functions. Thus larger virtuality causes a large EMC effect and presumably large modifications of the electromagnetic form factors.  Ref.~\cite{CiofidegliAtti:2007ork}  computed ratios of nuclear structure functions to free nucleon structure functions at a value of Bjorken $x=1/2$, where the effects of Fermi motion are small. It was also suggested that deep inelastic processes might be more effective than quasi-elastic processes for determining the medium modifications of form factors. This was because in the former higher momentum components are probed.
    
The idea of PLC suppression was taken up in Ref.~\cite{Kim:2022lng} in a calculation that used an AdS/QCD motivated light-front quark-diquark model to compute the nuclear structure functions. The related modifications of nucleon radii were also examined. The present paper presents a more detailed study of medium-modified form factors at all values of momentum transfer.
   
We summarize the work of Ref.~\cite{Kim:2022lng}. Motivated by the the QCD physics of color transparency \cite{Frankfurt:1985cv,Jennings:1993hw}, Ref.~\cite{Kim:2022lng} treated the infinite number of quark-gluon configurations of the nucleon as two configurations: a large-sized, blob-like configuration (BLC), consisting of complicated configurations of many quarks and gluons, and a small-sized, PLC consisting of 3 quarks. The BLC can be thought of as an object larger in average size than a nucleon, and the PLC is meant to represent a three-quark system of small size that is responsible for the high-$x$ behavior of the distribution function. This is because the smaller the number of quarks, the more likely one can carry a large momentum fraction.  Recent support for the idea that configurations of a proton with large values of $x$ have smaller than average size comes from an an analysis hadron production at backward rapidities and from an  analysis of jet production data in proton- and deuteron-nucleus collisions at RHIC and the LHC~\cite{PhysRevC.93.011902,PhysRevD.98.071502}.

When placed in a nucleus, the BLC feels the usual nuclear attraction, thus its energy decreases. The PLC feels far less nuclear-attraction by virtue of color screening \cite{Frankfurt:1994hf}, in which the effects of gluons emitted by small-sized configurations are cancelled in low-momentum transfer processes. The nuclear attraction increases the energy difference between the BLC and the PLC, therefore reducing the PLC probability~\cite{Frankfurt:1985cv}. Reducing the PLC probability in the nucleus reduces the quark momenta, in qualitative agreement with the EMC effect. Working out the consequences of the BLC-PLC model enables the connection between the EMC effect and virtuality to be clarified.

The Hamiltonian  for a free nucleon  in the two-component model is given by

\begin{equation}
    H_0 = \left[ \begin{array}{cc}
            E_B & V \\
            V & E_P \end{array} \right],
\end{equation}
    
\noindent where subscripts $P$ and $B$ represent the PLC and BLC respectively.
We define the energy difference between the PLC and the BLC to be $\D = E_P - E_B$.
The hard-interaction potential, $V$, connects the two
components, causing the eigenstates of $H_0$ to be  $|N\ra$ and $|X\ra$  rather
than $|B\ra$ and $|P\ra$. 
The normalized eigenstates are given by 
\begin{align} \label{nuc0}
    |N\ra = \frac{1}{\sqrt{1+\epsilon^2}}(|B\ra  +\epsilon|P\ra),  \notag\\
    |X\ra = \frac{1}{\sqrt{1+\epsilon^2}}(\e|B\ra -|P\ra),
\end{align} 

\noindent where $\epsilon = -2V/(\Delta + \bar\Delta)$ and $\bar\Delta\equiv\sqrt{\Delta^2+4V^2}$.
The notation $|X\ra$ is used to denote the orthogonal excited  state. The probability of the PLC, $P_{PLC}$, is then
$ 
    P_{PLC}= \frac{\epsilon^2}{1 + \epsilon^2}.
$


Now suppose the nucleon is bound to a nucleus. The nucleon feels an attractive, flavor-dependent, nuclear potential acting only on the BLC, so  the Hamiltonian is given by 
 \bea H= \left[ \begin{array}{cc}
E_B-|U_{N}| & V \\
V & E_P \end{array} \right],
 \eea
 
\noindent where $U_{N}$ is the attractive nuclear potential and the subscript $N$ is the flavor of the nucleon, proton or neutron. These interactions  increase the energy difference between the bare BLC and PLC states, and thereby decrease the PLC probability.


The medium-modified nucleon wave function
$|\tilde{N}\ra$ is  
\begin{equation} 
    |\tilde{N}\ra = \frac{1}{\sqrt{1+\tilde{\epsilon}^2_N}}(|B\ra  +\tilde{\epsilon}_N|P\ra),  
\end{equation}
with
$
\tilde{\e}_{N} \approx \e(1 - {|U_{N}|\over \bar\D}) \label{epsilon_mod}
$
 and 
the probability of the modified PLC for the nucleon, $\tilde{P}_{PLC}$, is  given by 




\begin{equation} \label{prob_mod}
 \tilde{P}_{PLC}= P_{PLC} \left(  1 - \frac{2|U_{N}|}{\bar{\D} (1 + \epsilon^2)  } \right).
\end{equation}.

\noindent The next step is to relate $  U_{N}$  to the fractional  virtuality $ \Cv_N=\Cv/M^2$, which is done in Ref. \cite{CiofidegliAtti:2007ork}. 
 The result is  
\bea \label{cv}
\Cv_{N} = {-2|U_{N}|\over M}, 
\eea
in which $U$ and $\Cv_n$ are  average quantities. 

Ref.~\cite{Kim:2022lng} introduced a  model of the EMC effect using LFHQCD nucleon form factors \cite{sufian_analysis_2017} as a starting point. In that work  the SU(6) spin-flavor symmetric quark model was used to calculate the effective charges of positive and negative helicity protons and neutrons. These effective charges were then used to obtain expressions for the nucleon form factors. The key idea is that the nuclear medium will affect the probabilities of finding a spin up or down quark $q$ in a proton or neutron, breaking SU(6) symmetry; thus, the effective charges are ultimately modified as well. 

We next consider free nucleon form factors and  PDFs from LFHQCD using Ref.\cite{sufian_analysis_2017}. In their work the form factor for an arbitrary twist-$\tau$ (number of partons) hadron is expressed as
\begin{equation} \label{hff}
    F_{\tau}(Q^2) ={1\over (1+{Q^2\over M^2_0}) (1+{Q^2\over M^2_1})\cdots (1+{Q^2\over M^2_{\tau-2}})}, 
   \end{equation}
a product of poles along the vector meson Regge radial trajectory in terms of the $\r$  vector meson mass, $M_\rho$, and its radial excitations of mass, $M_n$, with $M_n^2= 2M_\r^2(n+1/2)$.

The use of SU(6) symmetry lead  Ref. \cite{sufian_analysis_2017}  to obtain
\begin{equation} \label{f1p_nomod}
    F_1^p(Q^2)  = F_{\tau=3}(Q^2) 
\end{equation}

\begin{equation} \label{f1n_nomod}
    F_1^n(Q^2)  = -\frac{1}{2}F_{\tau=3}(Q^2)  + \frac{1}{2} F_{\tau = 4}(Q^2).
\end{equation}

\noindent One can obtain the up ($u$) and down ($d$) valence PDFs of the free proton and neutron by using a flavor decomposition of nucleon form factors \cite{cates_flavor_2011}, charge (isospin) symmetry~\cite{Miller:2006tv}, and by writing the form factors for quark flavor $q$, $F^q$, in terms of the valence GPD $H^q_v (x,t)$,


\begin{equation} \label{flavordecomp_isospin}
   (F_1^u)^p = 2F_1^p + F_1^n, \,\,\ (F_1^d)^p = F_1^p + 2F_1^n.
\end{equation}

\noindent Then  \cite{deTeramond:2018ecg},
\begin{equation} \label{flavorff}
    (F^q_1)^N = \int_0^1 dx \, H^q_v (x,t) = \int_0^1 dx \, q_v^N(x) e^{t f(x)},
\end{equation}
where $(F_1^q)^N$ is the $F_1$ flavor form factor for quark $q$ in nucleon $N$, $q_v^N(x)$ is the valence PDF for quark flavor $q$ in nucleon $N$, and $f(x)$ is the profile function. 

Furthermore, Ref. \cite{de_teramond_gaugegravity_2010} recast Eq. (\ref{hff}) in terms of an Euler Beta Function and determined what the PDF is for $(F^q_1)^N = F_\tau$. The corresponding PDF for $F_\tau$, referred to as $q_\tau(x)$, is normalized to unity and is given by,

\bea  &
q_\tau(x) =\frac{\Gamma \left(\tau -\frac{1}{2}\right)}{\sqrt{\pi } \Gamma (\tau -1)}\big(1- w(x)\big)^{\tau-2}\, w(x)^{- \half}\, w'(x), \,\,\, \,\,\, \label{qt}
  \eea
 with $ w(x) = x^{1-x} e^{-a (1-x)^2} $,
 where the flavor-independent parameter $a = 0.531 \pm 0.037$. Using Eqs. (\ref{f1p_nomod}, \ref{f1n_nomod}, \ref{flavordecomp_isospin}, \ref{flavorff}) one obtains
 \cite{deTeramond:2018ecg} the valence $u$ and $d$ proton quark distributions at the matching scale between LFHQCD and pQCD, $\m_0=1.06\pm 0.15 $ GeV. The size of the EMC effect is approximately independent of the factorization scale~\cite{Geesaman:1995yd,PRNorton_2003,Hen:2013oha,Rith:2014tma,Malace:2014uea,Hen:2016kwk}.
\bea &
u_v^p(x) ={3\over 2}q_3(x) +{1\over 2} q_4(x),\,\,
d_v^p(x)=q_4(x) \label{qdist}.
\eea

\noindent The neutron valence PDFs are obtained by using  isospin symmetry. The square of the proton and neutron wave functions are characterized by, $\Psi_p^2 \sim \frac{3}{2}q_3(x) + \frac{3}{2}q_4(x)$ and $\Psi_n^2 \sim \frac{3}{2}q_3(x) + \frac{3}{2}q_4(x)$, which are related to the elastic form factor $F_1^{N}$. 

The elastic form factors in the LFHQCD model, \eq{hff}, fall asymptotically as $1/Q^{2\t}$, and the slope of form factors as $Q^2=0$ increases with increasing values of  $\t$. Thus an increase in the value of $\t$ corresponds to an increase in effective size. Because $\tau$ represents  the number of constituents in a given Fock component of the hadron,  the function $q_3$ is naturally associated with the a three quark PLC system and $q_4$ with the BLC. This association is also consistent with the discussion regarding the PLC dominating at high-$x$ as can be seen by analyzing $g(x) \equiv  q_3/q_4$, which approaches infinity as $x$ nears 1. See Ref.~\cite{Kim:2022lng} for a plot. The exact expression for $g(x)$ is obtained from \eq{qt}. A qualitatively accurate representation of $g(x)$ for values of $x$ between 0.3 and 0.7 is given by $g(x)\approx 1/(1-x)^{4/3}.$ Lastly, the expressions for the wave function tell us that  the probability of the PLC, $P_{PLC}$, for both nucleons is equal to 1/2, thus $\epsilon^2 = 1$.

Ref.~\cite{Kim:2022lng} used holographic QCD to obtain
modified nucleon PDFs by using  the consequences of nuclear medium breaking  of SU(6) symmetry. 
Motivated by neutron-proton dominance in SRCs, we expect the nuclear potential to depend on whether one introduces a proton or neutron into the nucleus. For example, we expect the proton to feel a stronger attraction to a nucleus if there is an abundance of neutrons and vice versa. We apply medium effects by introducing two free
positively-valued  parameters that depend on mass and atomic numbers $A$ and $Z$,  $\delta r_p(A,Z)$ and $\delta r_n(A,Z)$. The $A$ and $Z$ dependencies in $\delta r_p$ and $\delta r_p$ are dropped to simplify the notation, but  are implied. The signs in front of $\delta r_p$ and $\delta r_n$ are motivated by the suppression of the PLC from the two-component model. The above parameterization leads to a suppression of the PLC, $q_3(x)$. 
The net result was
\begin{equation} \label{f1p_mod}
    \tilde{F}_1^p(Q^2)  = (1 - \delta r_p)F_{\tau=3}(Q^2)  + \delta r_p F_{\tau=4}(Q^2),
\end{equation}
and
\begin{equation} \label{f1n_mod}
     \tilde{F}_1^n(Q^2)  = -\left(\frac{1}{2} + \delta r_n\right)(F_{\tau=3}(Q^2)  -  F_{\tau = 4}(Q^2)).
\end{equation}
The Pauli form factor $F_2$ is not modified in the model of \cite{Kim:2022lng}.

The essence of our procedure is the dynamical connection between the medium-modifications, $\d r_{p,n}$ and the underlying dynamics of the nucleon in the medium as represented by the virtuality. The same model that describes the EMC effect is used to compute the medium modifications of the form factors. 
This is different than    the work of ~\cite{Xing:2023uhj} which  obtained huge suppression of the form factors. That work did not present a model of the nucleon wave function in the medium. Instead they utilized the SRC model from Ref. \cite{CLAS:2019vsb}, providing them with the ratio of SRC nucleons to deuteron structure functions, and a light-front quark-diquark AdS/QCD model \cite{Maji:2016yqo} that is different from LFHQCD, which gives them formulae connecting form factors and structure functions. They multiplied the ratio obtained from the SRC model by free proton and neutron structure functions from their AdS/QCD model. This gave them their results for the structure functions of SRC nucleons. Then, they modified the parameters of their AdS/QCD model to match the shape of the SRC structure functions, and used those parameters to approximate the SRC form factors in the AdS/QCD model.

Following the lengthy introduction, we  now  are in position  to study form factor modifications over a wide range of values of $Q^2$.
The medium modification ratios can now be obtained using \eq{f1p_mod} and \eq{f1p_nomod}, with the result:
\bea \label{f1mod}
{ \tilde{F}_1^p(Q^2)\over {F}_1^p(Q^2)}=(1-\d r_p) +{\d r_p\over 1+{Q^2\over 5M_\r^2}} \\
{ \tilde{F}_1^n(Q^2)\over {F}_1^n(Q^2)}= {1+2\d r_n}
\eea

The values of $\d r_{N}$ were obtained by describing EMC ratios and are given in Table I. These values are to be considered as averages over the nucleus. 
The medium modifications grow with increasing virtuality, which corresponds to increasing values of missing momentum.

The relation between $\d r_{N}$ and the virtuality is given by 
$
    \delta r_{N} = \frac{1}{4} \frac{|U_{N}|}{\bar{\D}}=- \frac{M}{8} \frac{ \Cv_{N}}{\bar{\Delta}}
$
 from Eq. (\ref{cv}).   
The quantity  $\delta r_{N}$ is proportional to the nuclear potential $|U_{N}|$, which is also proportional to the virtuality. We can determine $|U_{N}|$ from our fitted $\delta r_{N}$ values, and noting that the energy difference between the ground state and the first excited state in this model is $2V$ which is at least 500 MeV, the energy of the Roper Resonance. For $A > 4$, the values of  $\delta r_{N}$  range from 0.044 to 0.078, thus $ 80 \,\, \mbox{MeV} \leq |U_{N}| \leq 156 \,\, \mbox{MeV}$. We decompose the nuclear potential into mean field and SRC contributions. This is much larger than the shell model potential at the center of the nucleus which is about 
 50 MeV for the absolute value of the mean field \cite{Lilley:2009zz}. These values of $|U_{N}|$  are consistent with the finding that both mean-field  and SRC effects contribute to the average virtuality, roughly in the ratio $1:2$~\cite{CiofidegliAtti:2007ork}.
Indeed both effects must be involved, as noticed long ago~\cite{Hen:2013oha}. The fundamental forces that bind nuclei involve two and three nucleons. The mean-field is obtained by averaging the fundamental forces over  densities. This is only an approximation. There must be residual effects of the basic forces, those that cause short-range-correlations.

It is immediately apparent from the small values of $\d r_{N}$ in Table I that the effects of the modifications of the elastic electromagentic form factors are only a few percent. This is in contrast with the results of ~\cite{Xing:2023uhj}  which showed huge suppression factors. 
At this time, there are no experimental limits on modification of neutron radii, although a method has been suggested~\cite{PhysRevLett.103.082301}. Thus we focus on the medium modifications of the proton form factor, examining the limits at small and large values of $Q^2$. We find that the ratio of the medium-modified mean square radii to the free radii, $\tilde R_{p}^2/R_{p}^2$, from the $F_1$ form factor is
\bea
\tilde R_{p}^2/R_{p}^2= 1 + {3\over 20} \d r_p, \label{rm}
\eea 
a truly tiny effect.
The limit of large values of $Q^2$ is simply
\bea
{ \tilde{F}_1^p(Q^2)\over {F}_1^p(Q^2)}=(1-\d r_p) \label{fm}
,\eea
which is no more than a few percent.

\begin{table}[h]
    \centering
    \begin{tabular} 
    {|P{1.5cm}|P{2.2cm}|P{2.2cm}|}
    \hline
    \multicolumn{1}{|c}{Nucleus} & \multicolumn{2}{|c|}{Ref.~\cite{Kim:2022lng}}  \\ 
    \cline{2-3}
    \rule{0pt}{3ex}
    & $\delta r_p$ & $\delta r_n$    \\
    \hline
    \rule{0pt}{3ex} 
    $^2$H & 0.010 $\pm$ 0.003 & 0.010 $\pm$ 0.003  \\
    \hline
    \rule{0pt}{3ex} 
    $^3$He & 0.031 $\pm$ 0.003 & 0.061 $\pm$ 0.006    \\
    \hline
    \rule{0pt}{3ex} 
    $^3$H & 0.032 $\pm$ 0.006 & 0.016 $\pm$ 0.003   \\
    \hline
    \rule{0pt}{3ex} 
    $^4$He & 0.040 $\pm$ 0.004 & 0.040 $\pm$ 0.004     \\
    \hline
    \rule{0pt}{3ex} 
    $^9$Be & 0.044 $\pm$ 0.004 & 0.035 $\pm$ 0.003     \\
    \hline
    \rule{0pt}{3ex} 
    $^{12}$C & 0.049 $\pm$ 0.003 & 0.049 $\pm$ 0.003    \\
    \hline
    \rule{0pt}{3ex} 
    $^{27}$Al & 0.057 $\pm$ 0.003 & 0.053 $\pm$ 0.003   \\
    \hline
    \rule{0pt}{3ex} 
    $^{56}$Fe & 0.074 $\pm$ 0.003 & 0.064 $\pm$ 0.003   \\
    \hline
    \rule{0pt}{3ex} 
    $^{63}$Cu & 0.052 $\pm$ 0.003 & 0.044 $\pm$ 0.003    \\
    \hline
    \rule{0pt}{3ex} 
    $^{197}$Au & 0.072 $\pm$ 0.004 & 0.048 $\pm$ 0.003    \\
    \hline
    \rule{0pt}{3ex} 
    $^{208}$Pb & 0.078 $\pm$ 0.005 & 0.051 $\pm$ 0.003  \\
    \hline
    \end{tabular}
    \caption{The $\delta r_p$ and $\delta r_n$ medium modifications from Ref.~\cite{Kim:2022lng}. }
    \label{tab:tabulated_delta}
\end{table}

The free and medium-modified elastic form factors $F_1$, as determined in~\cite{Kim:2022lng}  are shown in Fig.~\ref{Mmod} for a Pb nucleus where the EMC effect is largest, providing an upper limit within the model.  The influence of the nuclear medium  is shown to be very modest indeed,  
in accord with \eq{f1mod}  and produces effects that do not violate the results of  any existing measurements. 
This is because the probability of the excited state components are proportional to the square of the size of the EMC effect.  Since the EMC effect size is of order 10\%, the probability of the excited state components is only of order 1\%.

\begin{figure}[h]
  \centering
 \includegraphics[width=0.45\textwidth]{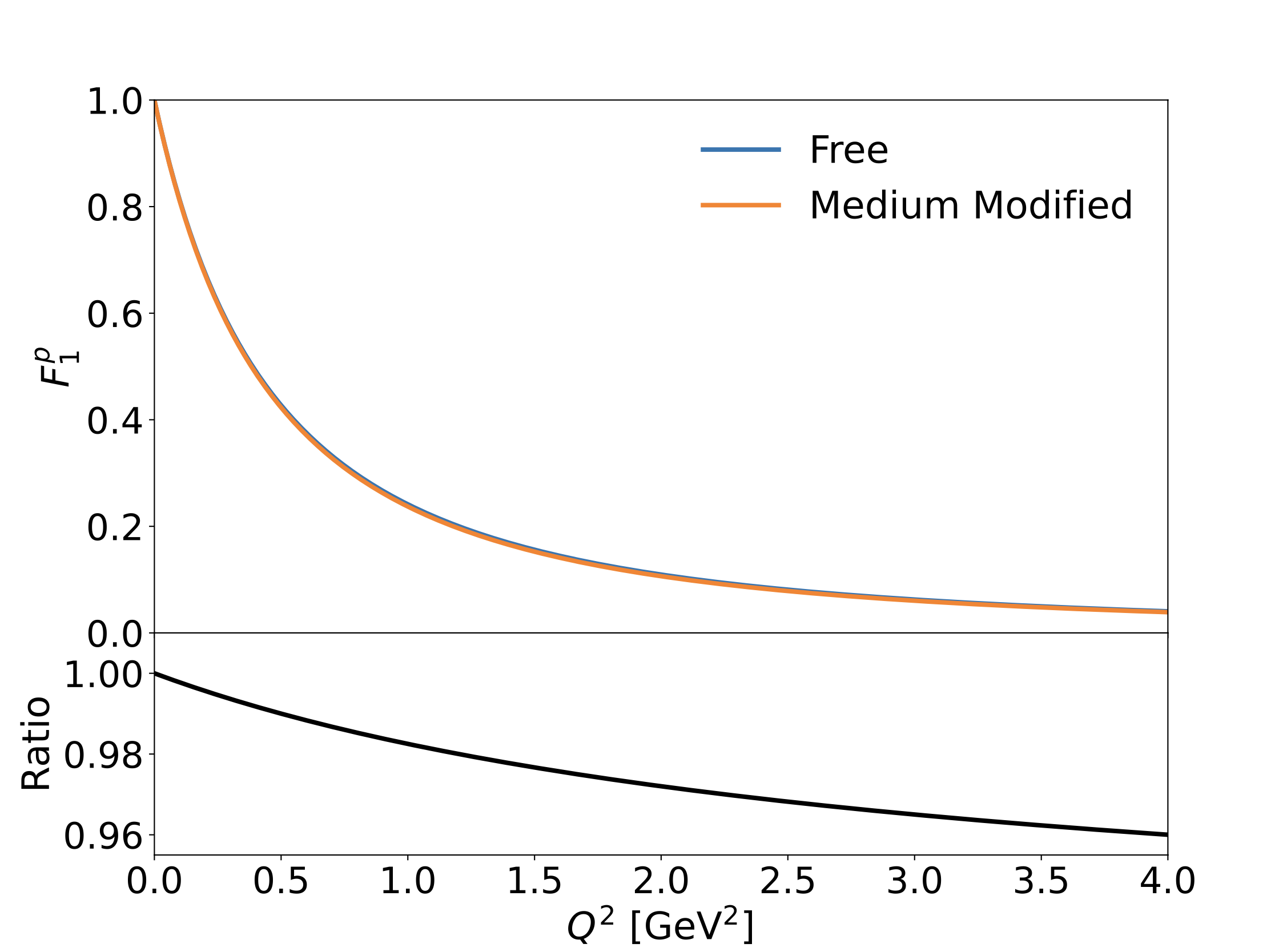}
  \hspace{0.3cm}
  \caption{(color online) (top)  Results for the $F_1$ elastic form factor of a free (blue) and medium modified (orange) proton in a Pb nucleus from Ref.~\cite{Kim:2022lng}. The bottom plot presents the ratio of the medium modified $F_1$ form factor to the free one.}
\label{Mmod}  \end{figure}

\begin{figure}[h]
  \centering
 \includegraphics[width=0.45\textwidth]{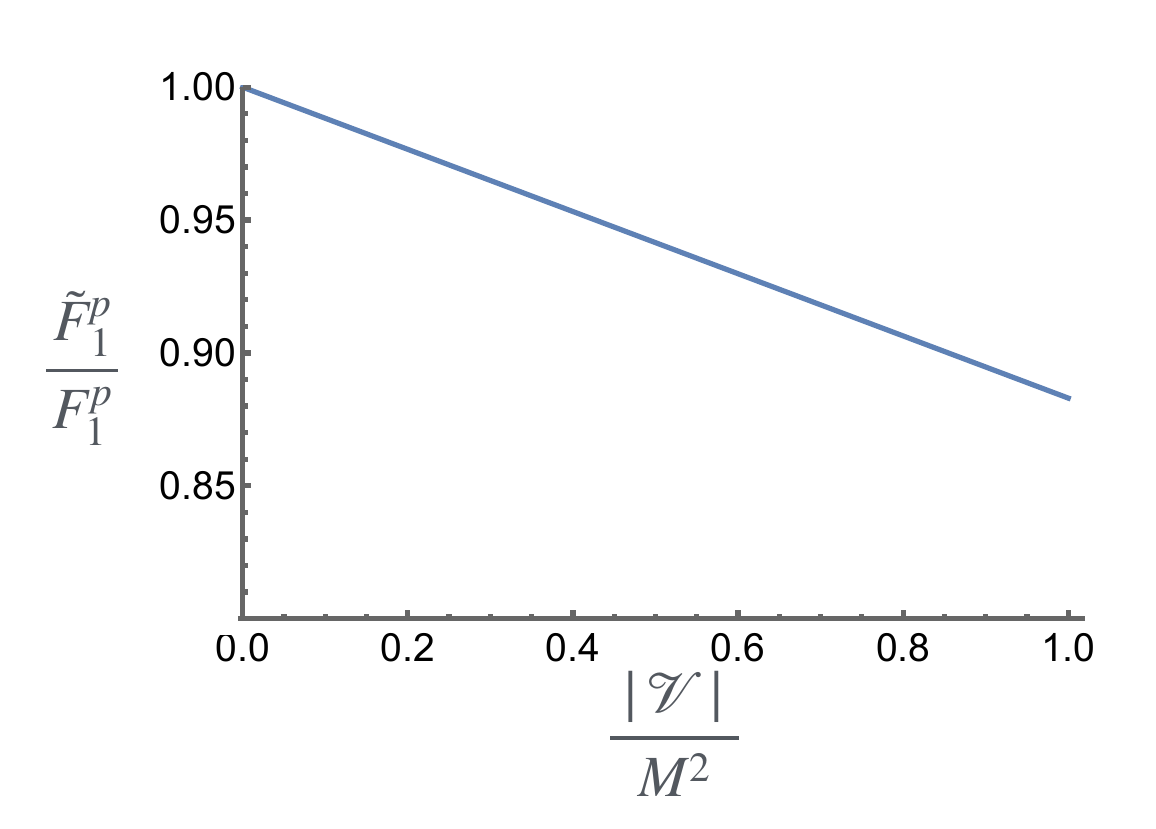}
  \hspace{0.3cm}
  \caption{(color online) (top) The ratio ${\tilde F^p_1\over F^p_1}$ for a fixed value of $Q^2= 5M_\rho^2=3$  GeV$^2$  as a function of virtuality in units of $M^2$. If the scattering is from a correlated pair, $\Cv\approx -2 \bfp^2$.}
\label{vdep}  \end{figure}

We also note that the size of the medium modification increases with virtuality, as shown in Fig.~\ref{vdep} which shows the ratio of medium-modified to free form factors is as a function of virtuality for a fixed value of $Q^2 =5M_\rho^2=3$  GeV$^2$.
Next turn to the medium modifications of the nucleon structure functions. 
Ref.~\cite{Kim:2022lng} found the EMC ratio, $R^{EMC}(A,Z)$, for nuclear structure function relative to deuterium, for a nucleus of mass and atomic numbers A and Z to
be
\bea&
R^{EMC}(A,Z)=     \frac{2}{A} \frac{F_2^A}{F_2^{d}} \nonumber\\&= \frac{2}{A} \frac{Z F_2^p + N F_2^n + \frac{5x}{3}(Z \delta r_p + N \delta r_n)(q_4 - q_3 )}{F_2^p + F_2^n + \frac{10x}{3}\delta r(^2H)(q_4 - q_3 )}.
\eea
We make a simple analysis to bring out the essential features of this result. Take $N=Z$ so that $\d r_p=\d r_n=\d r$ and also neglect the small effects of $\delta r(^2H)$. Then, with a bit of algebra, one finds that
\bea &
R^{EMC}(A,Z)
= 1+{5\over 9 } \d r {q_4(x)-q_3(x)\over q_4(x)+q_3(x)}\nonumber\\&\approx 1 +{5\over 9 } \d r (0.395-1.53 x), \label{rf}
\eea with the last expression being an accurate representation of the ratio for values $0.3\le x\le0.7.$ 
Comparing \eq{rm} and \eq{fm} with \eq{rf} it is evident that medium modification effects on elastic electromagnetic form factors and quark distribution functions are of about the same size. Thus, large modifications of the electromagnetic form factors are not required or even implied by an SRC-caused, virtuality-dependent EMC effect.

This paper shows that  one can construct a model in which the  medium modification of elastic electromagnetic form factors is about the same size   as the effects that cause the EMC effect even though  SRC are the origin. This contradicts the claims of Ref.~\cite{Xing:2023uhj}, so that the causal connection between SRC and the EMC effect is alive and well.
In any case, it would be very useful to obtain a direct experimental measurement of medium-modified form factors. It would be especially useful for such measurements to be made at kinematics where the nucleon virtuality is large, because all current measurements correspond to small values of the virtuality. A discussion of different experimental approaches is contained in~\cite{CLAS:2005ekq,PhysRevC.99.035205}, see also~\cite{Melnitchouk:1996vp}.

   {\bf Acknowledgments}
DNK and GAM would like to thank Guy F. de T\'eramond for pointing out an error in a previous version of this paper. The work of DNK and GAM  was supported by the U.S. Department of Energy Office of Science, Office of Nuclear Physics under Award No. DE- FG02-97ER41014. The research of MS was supported by the US Department of Energy Office of Science, Office of Nuclear Physics under Award No. DE-FG02-93ER40771. The work of LW was supported by the U.S. Department of Energy Office of Science, Office of Nuclear Physics under Award No. DE- FG02-96ER40960.

\maketitle     
\noindent

%

 
\end{document}